\documentclass[cameraready]{Interspeech}

\usepackage{times}
\usepackage{latexsym}
\usepackage[T1]{fontenc}
\usepackage[utf8]{inputenc}
\usepackage{microtype}
\usepackage{inconsolata}
\usepackage{amsmath}
\usepackage{amssymb}
\usepackage{url}
\usepackage{graphicx}
\usepackage{xcolor}
\usepackage{colortbl}
\usepackage{arydshln}       % must precede booktabs
\usepackage{booktabs}
\usepackage{multirow}
\usepackage{hyperref}
\usepackage{cleveref}
\usepackage{xcolor}

\title{Do Compact SSL Backbones Matter for Audio Deepfake Detection?\\
       A Controlled Study with RAPTOR}

\author[affiliation={1}, orcid=0009-0009-5915-6513]{Ajinkya}{Kulkarni}
\author[affiliation={2}]{Sandipana}{Dowerah}
\author[affiliation={3}]{Atharva}{Kulkarni}
\author[affiliation={2}]{Tanel}{Alumäe}
\author[affiliation={1}]{Mathew}{Magimai.-Doss}

\address{
    $^1$ Idiap Research Institute, Switzerland \\
    $^2$ Tallinn University of Technology, Estonia, $^3$ MBZUAI, UAE
}
\email{ajinkya.kulkarni@idiap.ch}

\keywords{Deepfake detection, Self-supervised learning, Test-time augmentation}

\begin{document}
\maketitle

% ---------------------------------------------------------------------------
%  ABSTRACT  (within 1000-char limit)
% ---------------------------------------------------------------------------
\begin{abstract}
Self-supervised learning (SSL) underpins modern audio deepfake detection, yet most prior work centers on a single large wav2vec2-XLSR backbone, leaving
compact under studied. We present RAPTOR, \textbf{R}epresentation \textbf{A}ware \textbf{P}airwise-gated \textbf{T}ransformer for \textbf{O}ut-of-domain \textbf{R}ecognition a controlled study of compact SSL backbones from the HuBERT and WavLM within a unified pairwise-gated fusion detector, evaluated across 14 cross-domain benchmarks. We show that multilingual HuBERT pre-training is the primary driver of cross-domain robustness, enabling 100M models to match larger and commercial systems. Beyond EER, we introduce a test-time augmentation protocol with perturbation-based aleatoric uncertainty to expose calibration differences invisible to standard metrics: WavLM variants exhibit overconfident miscalibration under perturbation, whereas iterative mHuBERT remains stable. These findings indicate that SSL pre-training trajectory, not model scale, drives reliable audio deepfake detection.
\end{abstract}

% =============================================================================
\section{Introduction}
% =============================================================================

Audio deepfakes have emerged as a serious threat to digital trust and
security \cite{kulkarni25b_interspeech}~\footnote{\scriptsize\url{https://www.europol.europa.eu/publications-events/publications/facing-reality-law-enforcement-and-challenge-of-deepfakes}}.
Recent advances in speech synthesis~\cite{ss1,ss2,ss3}, voice
conversion~\cite{ssvc1,ssvc2}, and neural audio generation have made highly
realistic synthetic speech widely accessible, enabling misuse in scenarios
such as fraud, impersonation, and
disinformation~\footnote{\scriptsize\url{https://www.ncsc.admin.ch/ncsc/de/home.html}}.
As a result, reliable audio deepfake detection has become a central research
problem in speech processing. Self-supervised learning (SSL) models have become
the \textit{de facto} feature extraction backbone for modern
detectors~\cite{dfarena,dfsuperb}, and subsequent improvements have largely focused on
the downstream classifier head, including graph-attention
architectures~\cite{jung2022aasist}, temporal convolution
modules~\cite{truong2024temporal,kulkarni25_interspeech}, and state-space
models~\cite{xlsrMamba,zhang2024audio,BiCrossMamba}. This design pattern has delivered
strong results on controlled benchmarks, yet recent large-scale evaluation
reveals that high in-domain performance does not reliably transfer to
out-of-domain conditions~\cite{dfarena,kulkarni2024generalization,tran25b_interspeech}, raising
a fundamental question about what truly drives detector robustness. This also reflects a broader shift in the literature from benchmark-centric binary detection toward robustness- and calibration-aware evaluation under distribution shift~\cite{muller24b_interspeech,pascu24_interspeech}.

A natural but underexplored hypothesis is that much of the cross-domain
behavior is determined not by the classifier, but by the SSL backbone itself.
Prior work has studied the contribution of individual SSL layers to deepfake
detection~\cite{layerwise}, finding that lower layers already carry
discriminative cues about synthesis artifacts. However, a controlled analysis
of how SSL \emph{pre-training trajectory} and \emph{backbone family} affect
downstream detection holding the classifier fixed remains absent.
This motivates our first research question: \textbf{\textit{RQ1.}} \textit{How does SSL
pre-training strategy, and in particular iterative multilingual refinement,
affect cross-domain audio deepfake detection performance?}

Orthogonal to pre-training strategy is the question of scale. Almost all
published high-performing systems \cite{dfarena,zhang2024audio,truong2024temporal,jung2022aasist,tran25b_interspeech,layerwise} rely on the 300M-parameter wav2vec2-XLSR
encoder \cite{xlsr}, while commercial systems may exceed 2B parameters. The practical case
for compact ${\sim}$100M SSL models lowers inference cost, ease the fine-tuning,
and deployment viability is clear, but whether such models can match their
larger counterparts on rigorous cross-domain benchmarks is an open question.
This defines our second question: \textbf{\textit{RQ2.}} \textit{Can compact ${\sim}$100M SSL
backbones deliver performance competitive with systems 5--20$\times$ larger,
including commercial deepfake detectors?}

Even when EER is used as the primary metric, it gives no signal about
\emph{how confidently} a model fails under distributional shift, a critical
consideration for real-world deployment where abstention or reliability scoring
may be required. Perturbation-based uncertainty estimation through test-time augmentation (TTA) has been explored in computer vision and, especially, medical imaging, where Monte Carlo predictions over transformed inputs have been used to quantify uncertainty and reveal cases that remain overconfident under distribution shift~\cite{wang2019tta,ayhan2020diagnostic,zhang2022memo}. Adapting this diagnostic to audio deepfake detection, where backbone
representations interact with acoustic perturbations in complex ways, motivates
our third question: \textbf{\textit{RQ3.}} \textit{Can TTA-derived aleatoric uncertainty
characterize SSL backbone confidence calibration in a way that standard EER cannot, and what
does this reveal about the relative robustness of compact SSL families?}

To address these three questions, we fix the downstream detection framework
to RAPTOR, a pairwise-gated hierarchical layer-fusion architecture used
consistently across all six backbones, and vary only the pretrained SSL encoder.
RAPTOR's primary role in this work is
as a controlled and interpretable evaluation setting. We conduct experiments
under two training protocols: single-dataset (ASVspoof~2019) and
Speech DF Arena \cite{dfarena} leaderboard and evaluate across 14 cross-domain
benchmarks. We introduce TTA with a perturbation-based
aleatoric uncertainty proxy ($U_{\text{ale}}$) to expose calibration differences
across SSL families. Our results show that compact iterative multilingual
pre-training outperforms not only other 100M systems but also 
larger commercial systems, and that $U_{\text{ale}}$ uncovers a exhibiting confidence-accuracy misalignment
overconfidence in WavLM variants that EER alone would miss.
% =============================================================================
\section{Method}
% =============================================================================

We perform a controlled comparison in which all systems share the same training
data, optimization setup, and downstream fusion architecture, varying only the
pretrained SSL encoder. The method comprises four components: compact SSL
backbone selection, the RAPTOR layer-fusion detector, consistency
regularization, and TTA-based uncertainty estimation.

% ---------------------------------------------------------------------------
\subsection{Compact SSL Backbone Families}
% ---------------------------------------------------------------------------

We study six compact SSL backbones of approximately 95--100M parameters
spanning two families and multiple pre-training trajectories
(Table~\ref{tab:ssl_backbones}). From the HuBERT family~\cite{hubert}, we
include HuBERT-Base (monolingual) and three multilingual mHuBERT
variants\footnote{\url{https://huggingface.co/utter-project/mHuBERT-147}}~\cite{mhubert} produced at successive stages of iterative
multilingual training: mHuBERT-Iter1, mHuBERT-Iter2, and mHuBERT-Final.
From the WavLM family~\cite{wavlm}, we include WavLM-Base and WavLM-Base+,
which share the same architecture but differ substantially in pre-training
data scale and diversity. Restricting all models to ${\sim}$100M parameters
isolates the effect of pre-training strategy and backbone family from raw
parameter count.

\begin{table}[!h]
\centering
\small
\caption{SSL backbone families and pre-training data. Lang.\ = languages.}
\label{tab:ssl_backbones}
\resizebox{\columnwidth}{!}{
\begin{tabular}{lll p{3.0cm}}
\hline
\textbf{Family} & \textbf{Model} & \textbf{Params} & \textbf{Pre-training data} \\
\hline
HuBERT & Base           & 95M & LibriSpeech 960\,h \\
HuBERT & mHuBERT-Iter1  & 95M & 90K\,h, 147 lang.\ (iter.\ 1) \\
HuBERT & mHuBERT-Iter2  & 95M & 90K\,h, 147 lang.\ (iter.\ 2) \\
HuBERT & mHuBERT-Final  & 95M & 90K\,h, 147 lang.\ (final) \\
WavLM  & Base           & 94M & LibriSpeech 960\,h \\
WavLM  & Base+          & 94M & 60K\,h + GigaSpeech + VoxPopuli \\
\hline
\end{tabular}}

\end{table}

% ---------------------------------------------------------------------------
\subsection{RAPTOR: Unified Layer-Fusion Detector}
% ---------------------------------------------------------------------------

As a fixed downstream framework used identically across all backbones, we
employ RAPTOR. Given input waveform $x$, the SSL encoder produces hidden
representations from $L$ transformer~\cite{transofrmer} layers,
\[
  \mathbf{H}=\bigl\{\mathbf{H}^{(1)},\ldots,\mathbf{H}^{(L)}\bigr\},
  \qquad \mathbf{H}^{(\ell)}\in\mathbb{R}^{T\times D},
\]
where $T$ is the sequence length and $D$ the feature dimension. RAPTOR then
fuses these layer representations through two learned gating stages before
attention pooling and binary classification (Fig.~\ref{fig:raptor}). \noindent\textbf{Pairwise gating:} Adjacent SSL layers $(\mathbf{H}^{(2p-1)},\mathbf{H}^{(2p)})$ are combined
by a time-dependent gate. A softmax over the concatenated pair produces
routing weights $\boldsymbol{\alpha}_p(t)=[\alpha_{p,1}(t),\alpha_{p,2}(t)]
\in\Delta^1$, yielding the fused frame representation:

\begin{figure}[t]
  \centering
  \includegraphics[width=\columnwidth]{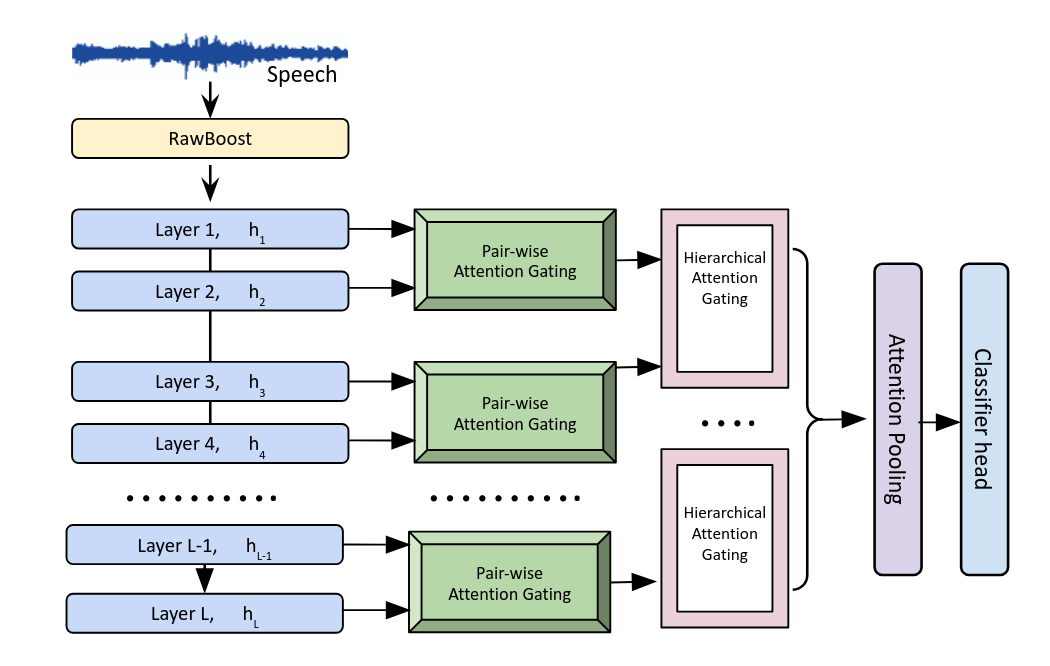}
  \caption{RAPTOR framework. SSL layer representations are
    progressively fused by pairwise and hierarchical softmax gates, followed
    by attention pooling and a binary classifier.}
  \label{fig:raptor}
  \vspace{-1em}
\end{figure}

\begin{equation}
  \tilde{\mathbf{h}}_p(t)
  = \alpha_{p,1}(t)\,\mathbf{h}_{2p-1}(t)
  + \alpha_{p,2}(t)\,\mathbf{h}_{2p}(t).
  \label{eq:pairgate}
\end{equation}
This allows the model to adaptively select artifact-relevant information from
neighboring SSL layers rather than relying on a fixed last-layer or uniform
average. A second hierarchical gating stage recursively fuses the pair-level
representations~$\tilde{\mathbf{h}}_p$ into a single utterance vector using
the same softmax routing mechanism, before attention pooling and the
classifier head. \newline

\noindent\textbf{Consistency regularization.}
Since both gating stages produce softmax distributions that lie on the
probability simplex, we apply a consistency regularization
term~\cite{miyato2019vat,tarvainen2017mean,xie2020uda} that encourages the
routing distributions to remain stable when the input is acoustically
perturbed. Given clean input $x$ and its RawBoost~\cite{rawboost1} augmented
view $\hat{x}$, we measure the symmetrized distributional discrepancy between
the corresponding gate activations using the Jensen--Shannon divergence, and
average this across all $M$ fusion modules and temporal positions. The final
training objective is:
\begin{equation}
  \mathcal{L}
  = \mathcal{L}_{\mathrm{cls}}(x,\hat{x},y)
  + \lambda\,\mathcal{L}_{\mathrm{cons}},
  \qquad \lambda = 0.25,
  \label{eq:loss}
\end{equation}
where $\mathcal{L}_{\mathrm{cls}}$ is class-weighted binary cross-entropy
over both views and $\mathcal{L}_{\mathrm{cons}}$ is the gating-distribution
consistency term. This encourages augmentation-invariant layer-selection
patterns and is particularly relevant in the compact-backbone setting, where
stable routing across SSL layers may matter more than raw parameter capacity.

% ---------------------------------------------------------------------------
\subsection{TTA-Based Uncertainty Estimation}
\label{sec:tta}
% ---------------------------------------------------------------------------

Standard EER is a point estimate that gives no information about whether
a model fails confidently or with appropriate uncertainty under distributional
shift. Uncertainty estimation through Monte Carlo dropout~\cite{gal2016dropout},
deep ensembles~\cite{lakshminarayanan2017}, and test-time
adaptation~\cite{wang2021tent} has demonstrated considerable attention in other
domains including medical imaging, autonomous driving, and natural language
processing as a diagnostic for identifying exhibiting confidence-accuracy misalignment predictions that standard
accuracy metrics do not reveal. We adapt this approach to audio deepfake
detection, where backbone representations interact with acoustic perturbations
in ways that EER alone cannot capture.

At test time, for each utterance $x$ we generate $K{=}3$ augmented views
using VoIP codec simulation, additive noise, and speed--pitch perturbation.
The detector produces a spoof posterior $p^{(k)}(x)$ for each view via sigmoid
on the output logit. The TTA mean posterior is:
\begin{equation}
  \bar{p}(x) = \frac{1}{K}\sum_{k=1}^{K}p^{(k)}(x).
  \label{eq:tta_mean}
\end{equation}

We then estimate the \emph{aleatoric uncertainty proxy} as the mean prediction
entropy across augmented views:

\begin{equation}
U_{\mathrm{ale}}(x)
=
\frac{1}{K}
\sum_{k=1}^{K}
\left(
-
\sum_{c=1}^{C}
p^{(k)}_{c}(x)\log p^{(k)}_{c}(x)
\right)
\label{eq:u_ale}
\end{equation}

\noindent where $H[p]=-p\log p-(1-p)\log(1-p)$. We interpret $U_{\mathrm{ale}}$ as an aleatoric-style proxy reflecting
sensitivity of the backbone's representations to acoustic input
perturbations~\cite{kendall2017uncertainties}, distinct from parameter-level
Bayesian uncertainty. TTA is evaluated in two complementary roles:
\textit{(a)} as an ensemble classifier using $\bar{p}(x)$ for EER computation
($\Delta$EER), and \textit{(b)} as a robustness diagnostic using $U_{\mathrm{ale}}$
to quantify per-backbone calibration under perturbation. In deployment settings,
$U_{\mathrm{ale}}$ can directly support reliability scoring and abstention
strategies: predictions with high $U_{\mathrm{ale}}$ signal that the backbone
representation is sensitive to acoustic conditions, warranting additional human
review or a fallback to a more conservative decision threshold.

% =============================================================================
\section{Experimental Setup}
% =============================================================================

% ---------------------------------------------------------------------------
\subsection{Datasets and Training Protocols}
\label{sec:datasets}
% ---------------------------------------------------------------------------

We consider two training protocols. \textbf{Protocol~1} trains exclusively
on ASVspoof~2019 \cite{todisco2019asvspoof2019futurehorizons}, enabling direct comparison with prior systems.
\textbf{Protocol~2} follows the Speech DF Arena recipe\footnote{\url{https://huggingface.co/Speech-Arena-2025/models}}\footnote{\url{https://huggingface.co/spaces/Speech-Arena-2025/Speech-DF-Arena}}~\cite{dfarena},
combining ASVspoof~2019\cite{todisco2019asvspoof2019futurehorizons}, ASVspoof~2024\cite{Wang2024_ASVspoof5}, CodecFake\cite{xie2024codecfake}, LibriSeVoc\cite{sun2023ai}, DFADD \cite{du2024dfadd},
CTRSVDD \cite{ctrsvdd}, SpoofCeleb\cite{spoofceleb}, MLAAD \cite{mlaad}, and EnvSDD \cite{envsdd}, increasing diversity in synthesis methods, codecs, and recording conditions. Offline augmentation using MUSAN~\cite{snyder2015musan} and room impulse response (RIR) simulation~\cite{ko2017study} expands each training utterance into five acoustic conditions (original, reverberation, speech, music, noise). Online RawBoost~\cite{rawboost1} augmentation is applied stochastically
per-batch during training. The TTA augmentations at inference (VoIP, noise,
perturbation) are distinct from both offline and online training augmentations.
Baseline systems Wav2Vec2-AASIST~\cite{tak2022automatic} and
Wav2Vec2-TCM~\cite{truong2024temporal} are trained under the same protocols;
DF-Arena 100M-V1 and DF-Arena 500M~\cite{dfarena} are included as external
reference points.

% ---------------------------------------------------------------------------
\subsection{Implementation Details}
% ---------------------------------------------------------------------------

All SSL backbone layers are fully fine-tuned jointly with the RAPTOR fusion
detector. Audio is resampled to 16\,kHz and cropped or zero-padded to 4\,s.
Models under Protocol~1 are trained for 50 epochs; Protocol~2 models are
trained for 100\,000 iterations. Both use the Adam optimizer with learning
rate $10^{-6}$, weight decay $10^{-4}$, and batch size 24 with full SSL fine-tuning. Consistency regularization weight is $\lambda{=}0.25$ (Eq.~\ref{eq:loss}). Model selection uses a held-out development set of the FoR dataset~\cite{reimao2019dataset}.

% ---------------------------------------------------------------------------
\subsection{Evaluation Protocols}
% ---------------------------------------------------------------------------

We follow the Speech DF Arena evaluation~\cite{dfarena} and report per-dataset
EER, average EER (mean across 14 sets), and pooled EER (single global threshold
from the combined score distribution). Pooled EER is the more stringent metric,
requiring consistent behavior across heterogeneous conditions under one shared
operating point. The 14 evaluation sets span ASVspoof (2019, 2021LA/DF,
2024)~\cite{todisco2019asvspoof2019futurehorizons,liu2023asvspoof,Wang2024_ASVspoof5},
ADD (2022, 2023; Track~1/3, Round~1/2)~\cite{yi2022add,yi2023add},
CodecFake~\cite{xie2024codecfake}, LibriSeVoc~\cite{sun2023ai},
SONAR~\cite{li2024sonarsyntheticaiaudiodetection},
FoR~\cite{reimao2019dataset}, DFADD~\cite{du2024dfadd}, and
ITW~\cite{muller2022does}.

% =============================================================================
\section{Results and Analysis}
% =============================================================================

% ------------------------------------------------------------------
\begin{table*}[!ht]
\centering
\caption{EER (\%) of RAPTOR and SOTA systems trained under Protocol~2
  (multi-dataset) across 14 cross-domain benchmarks. T\,=\,track,
  R\,=\,round, [P]\,=\,proprietary system (architecture and training details
  not publicly disclosed; included for reference only). \textbf{Bold} = best
  overall; \underline{underline} = best among 100M systems.}
\label{tab:eer}
\resizebox{\textwidth}{!}{%
\begin{tabular}{ll|c|cccc|c|c|cc|cc|c|c|c}
\toprule
& \textbf{System}
  & \textbf{ITW}
  & \multicolumn{4}{c|}{\textbf{ASVspoof}}
  & \textbf{FoR} & \textbf{CodecFake}
  & \multicolumn{2}{c|}{\textbf{ADD22}}
  & \multicolumn{2}{c|}{\textbf{ADD23}}
  & \textbf{DFADD} & \textbf{LibriSeVoc} & \textbf{SONAR} \\
& & & \textbf{19} & \textbf{21LA} & \textbf{21DF} & \textbf{24}
  & & & \textbf{T1} & \textbf{T3} & \textbf{R1} & \textbf{R2} & & & \\
\midrule
\parbox[t]{2mm}{\multirow{6}{*}{\rotatebox[origin=c]{90}{\textbf{SOTA systems}}}}
& W2V2-AASIST          & 2.12           & 1.88      &  6.01 & 1.78 & \textbf{11.31} &  3.80 & 13.35 & 22.76 & 3.62 & 12.43 & 17.44 & 0.23 & 0.08 &  1.07 \\
& W2V2-TCM             & 1.65           & 3.49      &  6.95 & 2.91 & 13.81 &  4.02 & 13.70 & 23.63 & 4.16 & 11.61 & 16.89 & 0.15 & 0.81 &  1.83 \\
& ResembleAI-2B [P]    & 3.94           & 1.32      &  \textbf{1.64} & 3.79 & 16.29 &  1.36 & 33.04 & 28.21 & 6.11 & 21.07 & 28.27 & 0.00 & 1.62 &  2.98 \\
& MoLEX [P]            & \textbf{0.03}  & \textbf{0.28} &  6.32 & 1.88 & 15.87 &  \textbf{0.17} & 32.40 & 31.93 & 3.66 & 11.11 & 19.02 & 6.46 & 3.30 &  \textbf{0.91} \\
& DF-Arena 500M        & 1.76 & 1.09    & 4.23 & 3.30 & 12.39 & 2.30 &\textbf{6.36}&23.98&\textbf{2.77}&\textbf{7.47}&\textbf{12.30}&\textbf{0.00}& 0.12 &1.90 \\
& DF-Arena 100M-V1     & 2.21 & 1.53    & 7.61 & 5.50 & 21.39 &  7.42 &  8.74 & 27.06 & 5.42 &  8.90 & 17.04 & 0.00 & 0.22 &  4.48 \\
\midrule
\parbox[t]{2mm}{\multirow{6}{*}{\rotatebox[origin=c]{90}{\textbf{RAPTOR (ours)}}}}
& HuBERT-Base          & 3.34 & 3.80 & 11.96 & 3.68 & 19.52 &  6.32 & 14.77 & 29.11 & 5.53 & 14.63 & 20.75 & 2.44 & 0.30 &  8.89 \\
& mHuBERT-Iter1        & 2.21 & 5.36 & 11.16 & 4.53 & 17.55 &  5.94 & \underline{13.34} & 24.31 & 4.75 & \underline{11.47} & \underline{16.10} & 0.40 & 0.67 &  9.09 \\
& mHuBERT-Iter2        & \underline{1.56} & \underline{2.70} & \underline{7.02} & 2.37 & \underline{16.01} & 3.14 & 14.04 & \textbf{\underline{22.06}} & 3.68 & 14.43 & 18.70 & 0.40 & \textbf{\underline{0.05}} & 3.44 \\
& mHuBERT-Final        & 1.62 & 3.21 &  9.16 & \textbf{\underline{1.83}} & 16.88 &  \underline{2.92} & 25.68 & 23.74 & \underline{3.56} & 14.89 & 20.52 & \underline{0.13} & 0.08 &  \underline{2.15} \\
& WavLM-Base           & 2.48 & 4.24 & 13.07 & 3.93 & 21.62 &  4.81 & 18.55 & 26.72 & 4.85 & 16.68 & 22.38 & 0.56 & 0.10 &  6.61 \\
& WavLM-Base+          & 2.24 & 2.84 &  8.66 & 3.59 & 18.45 &  4.59 & 13.79 & 26.38 & 6.61 & 13.95 & 19.59 & 0.68 & 0.18 &  8.59 \\
\bottomrule
\end{tabular}%
}
\vspace{-0.5em}
\end{table*}
% ------------------------------------------------------------------

% ------------------------------------------------------------------
\begin{table}[!t]
\centering
\small
\caption{Systems trained on ASVspoof~2019 only (Protocol~1), evaluated
  in-domain and on out-of-domain sets; DF Arena aggregate metrics are
  from multi-dataset training (Protocol~2). Avg\,EER\,=\,mean over 14 sets;
  Pooled\,EER\,=\,single global threshold across all 14 sets.}
\label{tab:asvtrain}
\resizebox{\linewidth}{!}{%
\begin{tabular}{l r ccc cc}
\toprule
\multirow{2}{*}{\textbf{System}}
  & \multirow{2}{*}{\textbf{Params}}
  & \multicolumn{3}{c}{\textbf{Protocol~1 (ASV19 train)}}
  & \multicolumn{2}{c}{\textbf{Protocol~2 (DF Arena Model)}} \\
\cmidrule(lr){3-5}\cmidrule(lr){6-7}
& & \textbf{ITW} & \textbf{FoR} & \textbf{ASV19}
& \textbf{Avg EER} & \textbf{Pooled EER} \\
\midrule
W2V2-AASIST    & 317M  & 11.19 &  7.46 & 0.22 &  6.99 & 12.46 \\
W2V2-TCM       & 319M  &  7.79 & 10.68 & 0.18 &  7.54 & 12.88 \\
ResembleAI-2B  & 2000M &  - &  - & - & 10.83 & 12.74 \\
MoLEX          &  376M &  - &  - & - &  9.51 & 12.40 \\
DF-Arena 500M  &  500M &  - &  - & - &  \textbf{5.78} & \textbf{10.88} \\
DF-Arena 100M-V1& 100M &  - &  - & - &  8.39 & 13.92 \\
\midrule
HuBERT-Base    & 100M  &  9.06 & 20.47 & 3.75 & 10.36 & 14.90 \\
mHuBERT-Iter1  & 100M  & 13.58 &  8.61 & 0.49 &  9.06 & 14.35 \\
mHuBERT-Iter2  & 100M  & 12.24 & 20.83 & 0.59 & \underline{7.83} & \underline{11.72} \\
mHuBERT-Final  & 100M  & 10.08 & 16.70 & 0.53 &  9.03 & 11.11 \\
WavLM-Base     & 100M  & 11.12 & 16.65 & 1.74 & 10.47 & 13.46 \\
WavLM-Base+    & 100M  &  9.27 & 12.30 & 1.15 &  9.30 & 13.97 \\
\midrule
\end{tabular}
}

\end{table}
% ------------------------------------------------------------------

% ---------------------------------------------------------------------------
\subsection{SSL Pre-Training Trajectory and Cross-Domain Robustness}
% ---------------------------------------------------------------------------

Table~\ref{tab:eer} reports per-dataset EER across all 14 benchmarks under
Protocol~2. The first observation is that multi-source training data
does not guarantee consistent cross-domain generalization. Even large-scale
proprietary systems exhibit EERs above 20\% on ASVspoof~2024, above 30\% on
CodecFake, and above 25\% on several ADD tracks, indicating that scale and
dataset breadth alone are insufficient to overcome sensitivity to unseen
synthesis methods, codec characteristics, and recording-condition mismatches.
ResembleAI-2B, despite its 2B-parameter architecture, reaches 33.04\% on
CodecFake and 28.27\% on ADD23-R2. MoLEX, while achieving 0.03\% on ITW,
degrades to 31.93\% on ADD22-T1. This finding implies that evaluation on any
single benchmark is insufficient to characterize detector robustness, and that
breadth of training data provides diminishing returns when the evaluation
distribution diverges substantially from training conditions.

Within the RAPTOR backbone family, mHuBERT-Iter2 achieves the most
consistently strong cross-domain performance: 1.56\% on ITW, 7.02\% on
ASVspoof~2021LA, 2.37\% on ASVspoof~2021DF, 16.01\% on ASVspoof~2024,
and 3.14\% on FoR. Crucially, the trajectory across mHuBERT checkpoints
reveals a broadly consistent effect of iterative multilingual pre-training.
HuBERT-Base (monolingual, 960h) achieves 3.34\% ITW and 11.96\% ASV21LA.
mHuBERT-Iter1 improves ITW to 2.21\% and FoR to 5.94\%. mHuBERT-Iter2
further reduces EER across the majority of benchmarks, achieving the best
average EER among all 100M systems (Table~\ref{tab:asvtrain}). Since the
downstream architecture and training setup are identical across all RAPTOR
systems, the performance differential is attributable specifically to the SSL
pre-training stage.

The progression breaks at mHuBERT-Final, which regresses substantially on
CodecFake (25.68\% vs.\ mHuBERT-Iter1's 13.34\% and mHuBERT-Iter2's
14.04\%). This non-monotonic behavior suggests that continued multilingual
pre-training beyond a certain stage may encode phonetic diversity at the
expense of low-level acoustic artifact sensitivity; precisely what
codec-based synthesis detection requires. The WavLM family shows a
different pattern: WavLM-Base+ outperforms WavLM-Base on most benchmarks
(8.66\% vs.\ 13.07\% on ASV21LA; 13.79\% vs.\ 18.55\% on CodecFake),
reflecting the benefit of larger pre-training data (60K vs.\ 960 hours),
but both WavLM variants remain weaker than mHuBERT-Iter2 in aggregate,
indicating that pre-training data volume alone does not substitute for
multilingual iterative refinement. \\

\noindent\textbf{\textit{RQ1:}} Iterative multilingual SSL pre-training is a first-order
factor in cross-domain audio deepfake detection robustness, independent of
downstream architecture. The controlled trajectory from HuBERT-Base to
mHuBERT-Iter2 demonstrates systematic improvement attributable to pre-training
strategy alone, while the regression at mHuBERT-Final reveals a
sensitivity--diversity trade-off that warrants further investigation.

% ---------------------------------------------------------------------------
\subsection{Compact 100M Systems vs.\ Large-Scale and Commercial Models}
% ---------------------------------------------------------------------------

Table~\ref{tab:asvtrain} compares compact RAPTOR variants against larger and
commercial systems under a common evaluation setting. Under Protocol~2,
mHuBERT-Iter2 achieves the best average EER among 100M systems (7.83\%),
while mHuBERT-Final achieves the best pooled EER among 100M systems (11.11\%).
This distinction matters because pooled EER measures consistency under a
single operating point across all 14 heterogeneous conditions.

Relative to 300M wav2vec2-XLSR systems, compact mHuBERT models are highly
competitive. mHuBERT-Iter2 improves pooled EER over W2V2-AASIST (12.46\%)
and W2V2-TCM (12.88\%) by 0.74 and 1.16 points respectively, using roughly
one-third of the parameters. mHuBERT-Final reduces pooled EER further to
11.11\%, reinforcing that compact multilingual SSL models can generalize
comparably to larger wav2vec2-XLSR backbones under cross-domain evaluation.
Both ResembleAI-2B (pooled EER 12.74\%) and MoLEX (12.40\%) are outperformed
by mHuBERT-Final on pooled EER. DF-Arena~500M remains the strongest overall
system (5.78\% average, 10.88\% pooled), demonstrating that scale continues
to help when paired with a purpose-built training setup; however, the compact
100M RAPTOR models clearly outperform the earlier DF-Arena~100M-V1 baseline
(8.39\% average, 13.92\% pooled).

Under Protocol~1 (ASVspoof~2019 training only), all systems degrade sharply
under domain shift. Although in-domain ASV19 EER is near zero for W2V2-TCM
(0.18\%), W2V2-AASIST (0.22\%), and the mHuBERT variants (0.49--0.59\%),
this does not transfer reliably to ITW and FoR. The 317M--319M wav2vec2
systems reach 7.79--11.19\% on ITW and 7.46--10.68\% on FoR, comparable to
the spread of the 100M compact systems.
This supports the conclusion that cross-domain robustness depends more on
SSL pre-training trajectory and training coverage than on backbone scale alone.

\noindent\textbf{\textit{RQ2:}} Compact 100M RAPTOR models do not surpass the strongest
purpose-built 500M system, but remain strongly competitive and outperform
larger 300M wav2vec2-XLSR systems and proprietary commercial detectors on
key cross-domain metrics, demonstrating that pre-training trajectory matters
more than scale alone.

% ---------------------------------------------------------------------------
\subsection{TTA-Based Uncertainty and Confidence Calibration}
\label{sec:tta_results}
% ---------------------------------------------------------------------------

\begin{table}[!t]
\centering
\small
\caption{$\Delta$EER (\%) $\downarrow$ and mean aleatoric uncertainty ($U_{\mathrm{ale}}$) $\uparrow$
  across $K{=}3$ TTA views (VoIP codec, additive noise, speed-pitch
  perturbation) for all systems on ITW, FoR, and ASV19, trained on \textit{Protocol 2}. $\Delta$EER is the
  TTA ensemble EER minus clean-inference EER; Systems exhibiting high $\Delta$EER alongside low
  $U_{\mathrm{ale}}$ demonstrate miscalibration under
  perturbation.}
\label{tab:delta_eer_uncertainty}
\resizebox{\linewidth}{!}{%
\begin{tabular}{l | rr | rr | rr}
\toprule
& \multicolumn{2}{c|}{\textbf{ITW}}
& \multicolumn{2}{c|}{\textbf{FoR}}
& \multicolumn{2}{c}{\textbf{ASV19}} \\
\cmidrule(lr){2-3}\cmidrule(lr){4-5}\cmidrule(lr){6-7}
\textbf{System}
  & $\Delta$\textbf{EER} & $U_{\mathrm{ale}}$
  & $\Delta$\textbf{EER} & $U_{\mathrm{ale}}$
  & $\Delta$\textbf{EER} & $U_{\mathrm{ale}}$ \\
\midrule
W2V2-TCM    & $+$1.70  & 0.299 & $+$44.79 & 0.247 & $+$1.06  & 0.214 \\
W2V2-AASIST & $+$1.83  & 0.227 & $+$45.03 & 0.263 & $+$1.57  & 0.124 \\
\midrule
HuBERT-Base    & $+$0.99  & \textbf{0.384} & \textbf{$+$42.51} & \textbf{0.418} & $+$0.48  & 0.289 \\
mHuBERT-Iter1  & $+$0.89  & 0.367 & $+$42.95 & 0.405 & $+$0.40  & \textbf{0.354} \\
mHuBERT-Iter2  & \textbf{$+$0.38}  & 0.321 & $+$45.72 & 0.321 & $+$1.66  & 0.277 \\
mHuBERT-Final  & \textbf{$+$0.38}  & 0.358 & $+$46.02 & 0.391 & \textbf{$-$0.18}  & 0.252 \\
WavLM-Base     & $+$13.88 & 0.274 & $+$44.13 & 0.265 & $+$16.57 & 0.190 \\
WavLM-Base+    & $+$13.14 & 0.214 & $+$44.74 & 0.168 & $+$9.82  & 0.141 \\
\bottomrule
\end{tabular}%
}
\vspace{-1em}
\end{table}

Table~\ref{tab:delta_eer_uncertainty} reveals a systematic pattern of
confidence miscalibration across SSL backbone families that standard EER
evaluation does not expose. The mHuBERT family exhibits small $\Delta$EER
values on ITW and ASV19, paired with moderate-to-high $U_{\mathrm{ale}}$.
mHuBERT-Iter1 shows $\Delta$EER\,=\,$+$0.89\% and $U_{\mathrm{ale}}$\,=\,0.367
on ITW, and $\Delta$EER\,=\,$+$0.40\% with $U_{\mathrm{ale}}$\,=\,0.354 on
ASV19. mHuBERT-Iter2 achieves $\Delta$EER\,=\,$+$0.38\% and
$U_{\mathrm{ale}}$\,=\,0.321 on ITW, and mHuBERT-Final shows a marginal
ensemble gain of $\Delta$EER\,=\,$-$0.18\% on ASV19 with
$U_{\mathrm{ale}}$\,=\,0.252. The comparatively higher $U_{\mathrm{ale}}$
across this family indicates that the backbone produces prediction entropy
that responds appropriately to acoustic perturbation -- a property consistent
with well-calibrated representations.

WavLM-Base and WavLM-Base+ exhibit a qualitatively different pattern that
constitutes the central finding of this analysis. WavLM-Base shows
$\Delta$EER\,=\,$+$13.88\% on ITW and $+$16.57\% on ASV19, while
$U_{\mathrm{ale}}$ remains at 0.274 and 0.190, among the lowest values in
the table. WavLM-Base+ shows $\Delta$EER\,=\,$+$13.14\% on ITW and $+$9.82\%
on ASV19, with $U_{\mathrm{ale}}$ of 0.214 and 0.141. This joint behavior
large EER degradation under perturbation alongside low aleatoric uncertainty
is the signature of overconfident miscalibration: the model produces
narrowly peaked posteriors across acoustically varied views, yet those
posteriors are inconsistent with the correct label. In deployment terms, a
detector with this property would not generate uncertainty signals sufficient
to trigger selective processing or human review, even under conditions where
its discrimination performance has substantially degraded.

W2V2-AASIST and W2V2-TCM, despite their 300M-parameter backbone, exhibit
moderate $\Delta$EER values on ITW ($+$1.83\% and $+$1.70\%) closer to the
mHuBERT family than to the WavLM family, with $U_{\mathrm{ale}}$ values of
0.227 and 0.299 respectively. This intermediate calibration profile offers a
further perspective on why 300M wav2vec2-XLSR variants do not close the
pooled-EER gap despite greater capacity.

On FoR, TTA produces EER increases above 42\% for all systems without
exception, $+$42.51\% for HuBERT-Base, $+$45.72\% for mHuBERT-Iter2,
$+$44.13\% for WavLM-Base, and $+$45.03\% for W2V2-AASIST. This uniform
degradation reveals a fundamental incompatibility between the
VoIP/noise/perturbation augmentation set and the specific acoustic
characteristics of FoR, and reinforces that $U_{\mathrm{ale}}$ and
$\Delta$EER must be evaluated jointly across multiple datasets to distinguish
backbone-level calibration properties from dataset-specific augmentation
effects. \\

\noindent\textbf{\textit{RQ3:}} TTA-based aleatoric uncertainty $U_{\mathrm{ale}}$ reveals
systematic overconfident miscalibration in WavLM variants; large $\Delta$EER
under acoustic perturbation alongside low $U_{\mathrm{ale}}$ not reflected by standard EER and constituting a distinct deployment risk beyond aggregate
discrimination metrics.

% ---------------------------------------------------------------------------
\section{Discussion}
% ---------------------------------------------------------------------------

The collective results yield several observations that extend beyond individual
benchmark results. \textbf{\textit{SSL pre-training trajectory governs
cross-domain transferability.}} The improvement from HuBERT-Base to
mHuBERT-Iter2 and the non-monotonic regression at mHuBERT-Final on
codec-based evaluation suggest that multilingual pre-training
selectively strengthens representations for cross-lingual acoustic
generalization up to a point, after which continued refinement may
over-specify toward language-specific features at the cost of
synthesis-artifact sensitivity. This is a distinct mechanism from data
scaling: mHuBERT-Iter2 and WavLM-Base+ have comparable pre-training data
volumes yet differ substantially in cross-domain EER and calibration behavior. \\

\noindent \textbf{\textit{Model scale does not substitute for pre-training quality.}}
mHuBERT-Iter2 at 100M parameters surpasses both 300M wav2vec2-XLSR systems
and the 2B-parameter ResembleAI commercial model on pooled EER, and
outperforms DF-Arena~100M-V1 by 2.20 pooled EER points despite a comparable
model size. These results support the hypothesis that representation
quality at the SSL pre-training stage is a more critical factor than downstream capacity or dataset aggregation. \\

\begin{figure}[!t]
  \centering
  \includegraphics[width=\columnwidth]{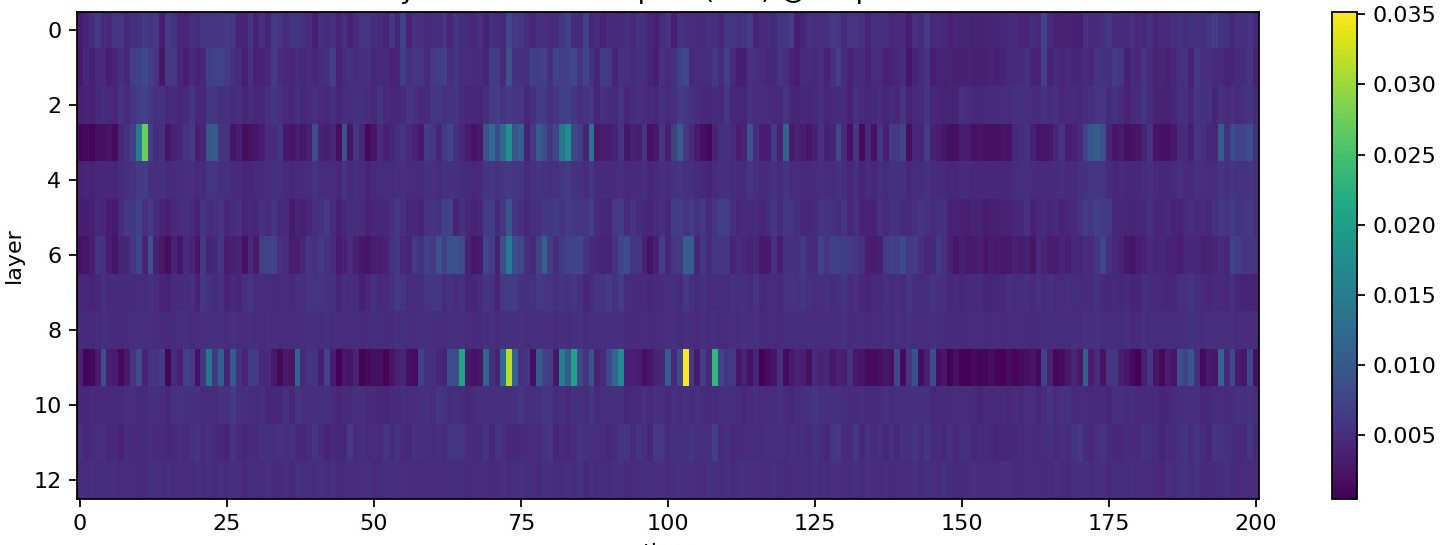}
  \caption{Pairwise gate maps $\alpha_{p,1}(t)$ for a spoofed utterance from ITW produced by mHuBERT-Iter2. The $x$-axis denotes time frames (50\,ms resolution), the $y$-axis the SSL layer-pair index ($p=1\ldots6$). Spoof utterances activate lower-to-middle layer pairs (indices 2--4) more strongly, suggesting synthesis artifacts concentrate at earlier stages of the SSL hierarchy.}
  \label{fig:inter}
\end{figure}

\noindent\textbf{\textit{Adversarial perturbations expose backbone-specific calibration
failure modes.}} The TTA results identify WavLM variants as exhibiting
overconfident miscalibration under acoustic perturbation, while mHuBERT
variants maintain stable $\Delta$EER with appropriate $U_{\mathrm{ale}}$.
This backbone-specific pattern indicates that the pre-training objective and
data composition affect not only discriminative accuracy but also the confidence
level of learned representations. WavLM's masked speech prediction
objective, combined with large-scale English pre-training, may produce
decision boundaries that are locally overconfident and sensitive to
distributional perturbations outside its training distribution. \textbf{\textit{Qualitative layer analysis:}} Pairwise gate maps in
Fig.~\ref{fig:inter} show that spoof utterances consistently activate
lower-to-middle SSL layer pairs more strongly than bona fide utterances,
consistent with prior layer-wise analysis~\cite{layerwise} and suggesting that
synthesis artifacts are preferentially captured at earlier stages of the SSL layer hierarchy. \\

%\noindent\textbf{\textit{Limitations:}} 

\noindent\textbf{\textit{Limitations:}} The TTA framework estimates aleatoric uncertainty via deterministic forward passes and does not provide epistemic
uncertainty, which requires weight-posterior inference \cite{uncer1}. Gate map analysis
remains qualitative; quantitative characterisation via layer-pair entropy and
gate consistency statistics is needed to substantiate artifact localisation. Future work will require to address epistemic uncertainty estimation through Bayesian approximation and ensemble methods, and investigate domain-adaptive TTA perturbation selection \cite{uncert2}.

% =============================================================================
\section{Conclusion}
% =============================================================================

We presented RAPTOR, a controlled study isolating the effect of SSL
pre-training trajectory on cross-domain audio deepfake detection under a fixed
pairwise-gated fusion framework. Our results establish three clear findings.
First, multilingual pre-training is a stronger predictor of
cross-domain robustness than backbone breadth or dataset scale: compact 100M
mHuBERT variants remain competitive with systems many times larger, including
purpose-built commercial detectors. Second, this advantage is not uniform
across pre-training stages, a non-monotonic regression at the final mHuBERT
checkpoint reveals a sensitivity--diversity trade-off that warrants further
investigation. Third, standard EER is insufficient to characterize deployment
reliability: TTA-based aleatoric uncertainty exposes systematic overconfident
miscalibration in WavLM variants that aggregate discrimination metrics cannot
detect. Collectively, these findings emphasize the role of pre-training strategy and calibration-aware evaluation in guiding system design.

Future work will extend the uncertainty framework to epistemic estimation
via Bayesian approximation and model ensembles. Gate map interpretability
will be quantified through layer-pair entropy and gate consistency statistics
to substantiate artifact localisation across SSL hierarchies.

% =============================================================================
\section{Generative AI Use Disclosure}
% =============================================================================

During the preparation of this work the authors used ChatGPT (GPT-4, OpenAI) in order to correct grammar and improve the fluency of some sentences. After using these services, the authors reviewed and edited the content as needed and take full responsibility for the content of the published article.

\bibliographystyle{IEEEtran}
\bibliography{mybib}

\end{document}